\documentclass[conference]{IEEEtran}
\usepackage{times}

\usepackage[numbers]{natbib}
\usepackage{multicol}
\usepackage[bookmarks=true]{hyperref}
\usepackage{amsthm}
\usepackage{booktabs} 
\usepackage{fancyhdr}
\usepackage{hyperref}
\usepackage{bm}
\usepackage{srcltx}
\usepackage{amssymb,latexsym,amsfonts,amsmath,amscd,amsthm} 
\usepackage{subfigure}
\usepackage{paralist}
 \usepackage{color}
 \usepackage{float}
\usepackage{verbatim}
\usepackage[lined,ruled,vlined]{algorithm2e}

\usepackage{placeins}

\usepackage{subfigure}

 \usepackage{acronym}
\providecommand{\abs}[1]{\left|#1\right|}
\providecommand{\norm}[1]{\lVert #1 \rVert}
\newcommand{\mix}{\mathrm{mix}}

\newcommand{\sink}{\mathsf{sink}}
\newcommand{\calC}{\mathcal{C}}
\newcommand{\calM}{\mathcal{M}}
\newcommand{\calA}{\mathcal{A}}
\newcommand{\calAP}{\mathcal{AP}}
\newcommand{\AEC}{\mathsf{AEC}}
\newcommand{\truev}{\mathsf{true}}
\newcommand{\falsev}{\mathsf{false}}
   \newcommand{\nat}{\mathbb{N}}   
    \newcommand{\hatH}{\hat{H}}   
    \renewcommand{\Pr}{\mathrm{Pr}}   

\theoremstyle{definition}
 \newtheorem{definition}{Definition}
 
\newtheorem{problem}{Problem}

\theoremstyle{plain}
\newtheorem{theorem}{Theorem}
 
 \newtheorem{lemma}{Lemma}
\theoremstyle{remark}

\acrodef{pdfa}[PDFA]{probabilistic deterministic finite-state automaton}
\acrodef{pfa}[PFA]{probabilistic finite-state automaton}
\acrodef{dba}[DBA]{deterministic B\"uchi automaton}
\acrodef{dra}[DRA]{deterministic Rabin automaton}

\acrodef{srs}[SRS]{stochastic reactive system}
\acrodef{ltl}[LTL]{linear temporal logic formula}
\acrodef{pltl}[PLTL]{probabilistic linear temporal logic formula}
\acrodef{mdp}[MDP]{Markov decision process}
\acrodef{pac-mdp}[PAC-MDP]{probably approximately correct Markov decision process}
\acrodef{rl}[RL]{reinforcement learning}
\acrodef{aec}[AEC]{accepting end component}
\usepackage{graphicx}
\graphicspath{{./figure/}}

\begin{document}
\title{Probably Approximately Correct MDP Learning and Control With
  Temporal Logic Constraints}

\author{Author Names Omitted for Anonymous Review. Paper-ID [add your ID here]}

\author{\authorblockN{Jie Fu and Ufuk Topcu}
  \authorblockA{Department of Electrical and Systems Engineering\\
    University of Pennsylvania\\
    Philadelphia, Pennsylvania 19104\\
    Email: jief,utopcu@seas.upenn.edu} }

\maketitle

\begin{abstract}

  We consider synthesis of control policies that maximize the
  probability of satisfying given temporal logic specifications in
  unknown, stochastic environments. We model the interaction between
  the system and its environment as a \ac{mdp} with initially unknown
  transition probabilities. The solution we develop builds on the
  so-called model-based probably approximately correct Markov decision
  process (PAC-MDP) methodology. The algorithm attains an
  $\varepsilon$-approximately optimal policy with probability
  $1-\delta$ using samples (i.e. observations), time and space that
  grow polynomially with the size of the \ac{mdp}, the size of the
  automaton expressing the temporal logic specification,
  $\frac{1}{\varepsilon}$, $\frac{1}{\delta}$ and a finite time
  horizon.  In this approach, the system maintains a model of the
  initially unknown \ac{mdp}, and constructs a product \ac{mdp} based
  on its \emph{learned model} and the specification automaton that
  expresses the temporal logic constraints. During execution, the
  policy is iteratively updated using observation of the transitions
  taken by the system. The iteration terminates in finitely many
  steps. With high probability, the resulting policy is such that, for
  any state, the difference between the probability of satisfying the
  specification under this policy and the optimal one is within a
  predefined bound.
\end{abstract}

\IEEEpeerreviewmaketitle

\section{Introduction}
Integrating model-based learning into control allows an agent to
complete its assigned mission by exploring its unknown environment,
using the gained knowledge to gradually approach an (approximately)
optimal policy. In this approach, learning and control complement each
other. For the controller to be effective, there is a need for correct
and sufficient knowledge of the system. Meanwhile, by exercising a
control policy, the agent obtains new percepts, which is then used in
learning to improve its model of the system.  In this paper, we
propose a method that extends model-based \ac{pac-mdp} reinforcement
learning to temporal logic constrained control for \emph{unknown},
\emph{stochastic} systems.

A stochastic system with incomplete knowledge can be modeled as an
\ac{mdp} in which the transition probabilities are unknown. Take a
robotic motion planning problem as an example. Different terrains
where the robot operates affect its dynamics in a way that, for the
same action of the robot, the probability distributions over the
arrived positions differ depending on the level and coarseness of
different grounds. The robot dynamics in an unknown terrain can be
modeled as an \ac{mdp} in which the transition probabilities are
unknown.  Acquiring such knowledge through observations of robot's
movement requires large, possibly infinite number of samples, which
is neither realizable nor affordable in practice. Alternatively, with
finite amount of samples, we may be able to approximate the
actual \ac{mdp} and reason about the optimality and correctness
(w.r.t. the underlying temporal logic specifications) of policies
synthesized using this approximation.

The thesis of this paper is to develop an algorithm that computational
efficiently updates the controller subject to temporal logic
constraints for an unknown \ac{mdp}. 
We extend the \ac{pac-mdp} method \cite{Kearns2002,Brafman2003} to
maximize the probability of satisfying a given temporal logic
specification in an \ac{mdp} with unknown transition probabilities. In
the proposed method, the agent maintains a model of the \ac{mdp}
learned from observations (transitions between different states
enabled by actions) and when the learning terminates, the learned
\ac{mdp} approximates the true \ac{mdp} to a specified degree, with a
pre-defined high probability. The algorithm balances exploration and
exploitation implicitly: Before the learning stops, either the current
policy is approximately optimal, or new information can be invoked by
exercising this policy. Finally, at convergence, the policy is ensured
to be \emph{approximately optimal}, and the time, space, and sample
complexity of achieving this policy is polynomial in the size of the
\ac{mdp}, in the size of the automaton expressing the temporal logic
specification and other quantities that measure the accuracy of, and
the confidence in, the learned \ac{mdp} with respect to the true one.

Existing results in temporal logic constrained verification and control
synthesis with unknown systems are mainly in two categories: The first
uses statistical model checking and hypothesis testing for Markov
chains \cite{legay2010statistical} and \ac{mdp}s
\cite{Henriques2012}. The second applies inference algorithms to
identify the unknown factors and adapt the controller with the
inferred model (a probabilistic automaton, or a two-player
deterministic game) of the system and its environment
\cite{Chen2012,FuAtAlCDC2013}. Statistical model checking for
\ac{mdp}s \cite{Henriques2012} relies on sampling of the trajectories
of  Markov chains induced from the underlying \ac{mdp} and 
policies to verify whether the probability of satisfying a
\emph{bounded} linear temporal logic constraint is greater than some
quantity for all admissible policies.  It is restricted to
\emph{bounded} linear temporal logic properties in order to make the
sampling and checking for paths computationally feasible. For linear
temporal logic specifications in general, computationally efficient
algorithm has not been developed.  Reference \cite{abs-1212-3873}
employs inference algorithms for deterministic probabilistic
finite-state automata to identify a subclass of \ac{mdp}s, namely,
deterministic \ac{mdp}s.  Yet, this method requires the data (the
state-action sequences in the \ac{mdp}s) to be independent and
identically distributed. Such an assumption cannot hold in the
paradigm where learning (exploration) and policy update (exploitation)
are carried out in parallel and at run time, simply because that the
controller/policy introduces sampling bias for observations of the
system. Reference \cite{Chen2012} applies stochastic automata learning
combined with probabilistic model checking for stochastic
systems. However, it requires an \emph{infinite amount} of experiences
for the model to be identified and the policy to be optimal, and may
not be affordable in practice.

We show that the extension of the \ac{pac-mdp} method to control synthesis
subject to temporal logic constraints shares many attractive features
with the original method: First, it applies to linear temporal logic specifications
and guarantees efficient convergence to an approximately optimal
policy within a \emph{finite time horizon} and the number of policy
updates is determined by the size of underlying \ac{mdp}, independent
from the specification. Second, it balances the exploration (for
improving the knowledge of the model) and exploitation (for maximizing
the probability of satisfying the specification) and does not require
the samples to be independent and identically distributed.


\section{Preliminaries}
\label{sec:prelim}


\begin{definition}
  A labeled \ac{mdp} is a tuple $ M= \langle Q, \Sigma, q_0,P, \calAP,
  L \rangle $ where $Q$ and $\Sigma$ are finite state and action sets.
  $q_0\in Q$ is the initial state.  The transition probability
  function $P: Q \times \Sigma \times Q \rightarrow [0,1]$ is defined
  such that $\sum_{q' \in Q} P(q,\sigma,q') \in \{0,1\}$ for any state
  $q\in Q$ and any action $\sigma \in \Sigma$.  $\calAP$ is a finite
  set of atomic propositions and $L: Q \rightarrow 2^{\calAP}$ is a
  labeling function which assigns to each state $q \in Q$ a set of
  atomic propositions $L(q)\subseteq \calAP$ that are valid at the
  state $q$. $L$ can be extended to state sequences in the usual way,
  i.e., $L(\rho_1\rho_2)=L(\rho_1)L(\rho_2)$ for $\rho_1,\rho_2\in
  Q^\ast$.
\end{definition}

The \emph{structure} of the labeled \ac{mdp} $M$ is the underlying
graph $\langle Q, \Sigma, E\rangle$ where $E \subseteq Q\times
\Sigma\times Q$ is the set of labeled edges. $(q,\sigma,q')\in E$ if
and only if $P(q,\sigma,q')\ne 0$. We say action $\sigma$ is
\emph{enabled} at $q$ if and only if there exists $q'\in Q$,
$(q,\sigma,q')\in E$. 

A \emph{deterministic policy}  $f: Q^\ast\rightarrow \Sigma$ is such
that given  $\rho = q_0\ldots q_n$, $f(\rho)=\sigma$ only if $\sigma$ is
enabled at $q_n$.

\subsection{A specification language}
We consider to use \ac{ltl} to specify a set of desired system
properties such as safety, liveness, persistence and stability.  A
formula in \ac{ltl} is built from a finite set of atomic propositions
$\calAP$, $\truev$, $\falsev$ and the Boolean and temporal connectives
$\land, \lor, \neg, \Rightarrow, \Leftrightarrow$ and $\square$
(always), $\mathcal{U}$ (until), $\lozenge$ (eventually), $\bigcirc$
(next). Given a \ac{ltl} formula $\varphi$ as the system
specification, one can always represent it by a \ac{dra}
$\calA_\varphi=\langle S,2^{\calAP}, T_s,I_s, \mathsf{Acc} \rangle$
where $S$ is a finite state set, $2^{\calAP}$ is the alphabet, $I_s\in
S$ is the initial state, and $T_s: S\times 2^{\calAP} \rightarrow S$
the transition function.  The acceptance condition $\mathsf{Acc}$ is a
set of tuples $\{(J_i, K_i)\mid i =0, 1,\ldots,m\}$ consisting of
subsets $J_i$ and $K_i$ of $S$. The run for an infinite word $w=
w[0]w[1]\ldots \in (2^{\calAP})^\omega$ is the infinite sequence of
states $s_0s_1\ldots \in S^\omega$ where $s_0=I_s$ and
$s_{i+1}=T_s(s_i, w[i])$. A run $\rho=s_0s_1\ldots $ is accepted in
$\calA_\varphi$ if there exists at least one pair $(J_i,K_i) \in
\mathsf{Acc}$ such that $\mathsf{Inf}(\rho)\cap J_i =\emptyset$ and
$\mathsf{Inf}(\rho)\cap K_i \ne \emptyset$ where $\mathsf{Inf}(\rho)$
is the set of states that appear infinitely often in $\rho$.

Given an \ac{mdp} and a \ac{ltl} specification $\varphi$, we aims to
maximize the probability of satisfying $\varphi$ from a given
state. Such an objective is \emph{quantitative}
\cite{rutten2004mathematical}.
\subsection{Policy synthesis in a known \ac{mdp}}
We now present a standard quantitative synthesis method in a known
\ac{mdp} with \ac{ltl} specifications, following from
\cite{Baire2004,rutten2004mathematical}.
\begin{definition} 
  Given an \ac{mdp} $M =\langle Q, \Sigma, P, q_0, \calAP, L \rangle$
  and the \ac{dra} $\mathcal{A}_\varphi= \langle S ,2^{\calAP}, T_s,
  I_s, \{(J_i,K_i)\mid i =1,\ldots, m\} \rangle$, the \emph{product
    \ac{mdp}} is
$
\mathcal{M}= M \ltimes \mathcal{A}_\varphi  =\langle V, \Sigma,
\Delta,v_0, \mathsf{Acc} \rangle
$,\\
with components defined as follows: $ V=Q\times S$ is the set of
states.  $\Sigma$ is the set of actions. The initial state is
$v_0=(q_0, s_0 ) $ where $s_0 =T_s(I_s,L(q_0))$.  $\Delta: V\times
\Sigma \times V\rightarrow [0,1]$ is the transition probability
function. Given $v= (q,s)$, $\sigma$, $v'=(q',s')$, let $\Delta
(v,\sigma,v')= P(q,\sigma,q')$ and $T_s(s, L(q'))=s'$.  The acceptance
condition is $\mathsf{Acc}=\{(\hat{J_i}, \hat{K_i}), i: =1,\ldots, m
\mid \hat{J_i} =Q\times J_i, \hat{K_i}=Q\times K_i \}$, which is
obtained by lifting of the set $J_i,K_i \subseteq S$ the
acceptance condition of $\calA_\varphi$ into $\calM$.
\end{definition}

A \emph{memoryless, deterministic policy} for a product \ac{mdp}
$\mathcal{M}=\langle V, \Sigma, \Delta,v_0, \mathsf{Acc} \rangle$ is a
function $f: V\rightarrow \Sigma$.  A memoryless policy $f$ in $\calM$
is in fact a finite-memory policy $f'$ in the underlying \ac{mdp}
$M$. Given a state $(q,s)\in V$, we can consider $s$ to be a memory
state, and define $f'(\rho)=f((q,s))$ where the run $\rho=q_0q_1\ldots
q_n$ satisfies $q_n=q$ and $T_s(I_s, L(\rho))=s$.

For the types of \ac{mdp}s, which are one-player stochastic games,
memoryless, deterministic policies in the product \ac{mdp} are
sufficient to achieve the quantitative temporal logic objectives
\cite{bianco1995model}. In this work, by policy, we refer to
memoryless, deterministic policy. In Section~\ref{sec:conclusion}, we
briefly discuss the extension of \ac{pac-mdp} method to two-player
stochastic games.

\begin{definition}[Markov chain induced by a policy]
  Given an \ac{mdp} $\mathcal{M}=\langle V, \Sigma,
  \Delta,v_0, \mathsf{Acc} \rangle$ and a policy $f :V \rightarrow
  \Sigma $, the \emph{Markov chain induced by policy $f$} is a tuple
  $\mathcal{M}^f = \langle V, \Sigma, \Delta^f, v_0,\mathsf{Acc} \rangle$
  where $\Delta^f(v,v')= \Delta(v,f(v),v')$.
\end{definition}

A \emph{path} in a Markov chain is a (finite or infinite) sequence of
states $x\in V^\ast$ (or $V^\omega$).  Given a Markov chain
$\mathcal{M}^f$, starting from the initial state $v_0$, the state
visited at the step $t$ is a random variable $X_t$. The probability of
reaching state $v'$ from state $v$ in one step, denoted $\Pr(X_{t+1}=
v' \mid X_t=v )$, equals $ \Delta^f(v,v')$. This is extended to a unique
measure $\Pr$ over a set of (infinite) paths of $\calM^f$,
$\Pr(v_0v_1\ldots v_n) =
\Pr(X_{n}=v_n \mid X_{n-1}=v_{n-1})\cdot
\Pr(v_0 v_1\ldots v_{n-1})$.


The following notations are used in the rest of the paper: For a
Markov chain $\calM^f$, let $h^{\le i}(v,X)$ (resp. $h^{i} (v,X)$) be
the probability of that a path starts from state $v$ and hits the set
$X$ for the first time within $i$ steps (resp. at the exact $i$-th
step). By definition, $h^{\le i} (v,X) = \sum_{k=0}^i h^i (q,X)$. In
addition, let $h (v,X) = \sum_{k=0}^\infty h^k(v,X)$, which is the
probability of a path that starts from state $v$ and enters the set
$X$ eventually. When multiple Markov chains are involved, we write
$h_{\calM^f}$ and $\Pr_{\calM^f}$ to distinguish the hitting
probability $h$ and the probability measure $\Pr$ in $\calM^f$.

\begin{definition}
  The \emph{end component} for the product \ac{mdp} $\mathcal{M}$
  denotes a pair $(W, f)$ where $W\subseteq V $ is non-empty and $f: W
  \rightarrow \Sigma$ is defined such that 
  \ for any $v\in W$, $\sum_{v'\in W}\Delta(v,
    f(v), v')=1$;
  and the induced directed graph $(W, \rightarrow_f)$ is strongly
    connected. Here, $v\rightarrow_f v'$ is an edge in the
    directed graph if $\Delta(v, f(v), v') >0$.
An \ac{aec} is an end component such that $W\cap \hat{J_i} =\emptyset$ and $W\cap \hat{K_i} \ne \emptyset$ for some $i \in \{1,\ldots, m\}$.
\end{definition}
Let the set of \ac{aec}s in $\calM$ be denoted $\AEC(\calM)$ and let
the set of \emph{accepting end states} be $\calC= \{v \mid \exists
(W,f)\in \AEC(\calM), v\in W\}$.  Due to the property of \ac{aec}s,
once we enter some state $v \in \calC$, we can find an \ac{aec}
$(W,f)$ such that $v\in W$, and initiate the policy $f$ such that for
some $i\in \{1,\ldots, m\}$, all states in $\hat{J_i}$ will be visited
only finite number of times and some state in $\hat{K_i} $ will be
visited infinitely often.  Given the structure of $\calM$, the set
$\AEC(\calM)$ can be computed by algorithms in
\cite{de1997formal,Chatterjee2012}. Therefore, given the system
\ac{mdp} $M$ and its specification automaton $\mathcal{A}_\varphi$, to
maximize the probability of satisfying the specification, we want to
synthesize a policy $f$ that maximizes the probability of hitting the
set of accepting end states $\calC$, and after hitting the set, a
policy in the accepting end component will be followed.

\subsection{Problem statement}
The synthesis method in Section \ref{sec:prelim} produces the optimal
policy for quantitative temporal logic objectives only if the \ac{mdp}
model is known.  However, in practice, such a knowledge of the
underlying \ac{mdp} may not be available. One example can be the
robotic motion planning in an unknown terrain.

Model-based reinforcement learning approach suggests the system learns
a model of the true \ac{mdp} on the run, and uses the knowledge to
iteratively updates the synthesized policy. Moreover, the learning and
policy update shall be efficient and eventually the policy converges
to one which meets a certain criterion of success. Tasked with
maximizing the probability of satisfying the specification, we define,
for a given policy, the state value in the product \ac{mdp} is
the probability satisfying the specification from that state onwards
and the optimal policy is the one that maximizes the state value for
each individual state in the product \ac{mdp}. The probability of
satisfying a \ac{ltl} specification is indeed the probability of
entering the set of accepting end states in the product \ac{mdp} (see
Section~\ref{sec:prelim}). We introduce the following definition.

\begin{definition}
  Let $\mathcal{M}$ be the product \ac{mdp}, $\AEC(\calM)$ be the set
  of accepting end components, and $f$ be a policy in
  $\mathcal{M}$. For each state $v\in V$, given a finite horizon $T\in
  \nat$, the $T$-step state value is $ U^f_{\cal M}(v, T)=h^{\le
    T}_{\calM^f}(v, \calC) $, where $\calC $ is the set of accepting
  end states obtained from $\AEC(\calM)$.  The \emph{optimal $T$-step
    state value} is $ U_{\cal M}^\ast(v,T)= \max_{f}\{ U_{\cal M}^f
  (v, T)\} $, and the \emph{optimal $T$-step policy} is $f^\ast_T=\arg
  \max_f \{ U_{\cal M}^f (v, T)\}$.  Similarly, We define the state
  value $U_{\calM}^f(v)=h_{\calM^f}(v, \calC) $. The \emph{optimal
    state value} is $ U_{\cal M}^\ast(v)= \max_{f}\{ U_{\cal M}^f
  (v)\}$ and the \emph{optimal policy} is $f^\ast = \arg\max_{f}\{
  U_{\cal M}^f (v)\}$.
\end{definition}
The definition of state-value (resp. $T$-step state value) above can
also be understood as the following: For a transition from state $v$
to $v'$, the reward is $0$ if neither $v$ or $v'$ is in $ \calC$ or if
$v\in \calC$ and $v'\in \calC$; the reward is $1$ if $v \notin \calC$
and $v'\in \calC$ and prior to visiting $v$, no state in $\calC$ has
been visited. Given a state $v$, its state value (resp. $T$-step state
value) for a given policy is the expectation on the eventually
(resp. $T$ steps) accumulated reward from $v$ under the policy.


We can now state the main problem of the paper.
\begin{problem}
\label{mainproblem}
Given an \ac{mdp} $ M=\langle Q,\Sigma, q_0, P, \calAP, L\rangle $
with unknown transition probability function $P$, and a \ac{ltl}
specification automaton $\calA_\varphi=\langle S , 2^{\calAP},
T_s,I_s, \mathsf{Acc}\rangle$, design an algorithm which with
probability at least $1-\delta$, outputs a policy $f:Q\times S
\rightarrow \Sigma$ such that for any state $(q, s)$, the $T$-step
state value of policy $f$ is $\varepsilon$-close to the optimal state
value in $\calM$, and the sample, space and time complexity required
for this algorithm is less than some polynomial in the relevant
quantities $(\abs{Q}, \abs{S}, \abs{\Sigma}, \frac{1}{\varepsilon},T,
\frac{1}{\delta})$.
  \end{problem} 
\section{Main result}
\label{sec:mainresult}
\subsection{Overview}
First we provide an overview of our solution to
Problem~\ref{mainproblem}. Assume that the system has full
observations over the state and action spaces, in the underlying
\ac{mdp} $M$, the set of states are partitioned into \emph{known} and
\emph{unknown} states (see
Definition~\ref{def:knownstates}). Informally, a state becomes known
if it has been visited sufficiently many times, which is determined by
some confidence level $1-\delta$ and a parameter $\epsilon$, the
number of states and the number of actions in $M$, and a finite-time
horizon $T$.

Since the true \ac{mdp} is unknown, we maintain and update a learned
\ac{mdp} $\overline{M}$, consequently $\overline{\calM}$.  Based on
the partition of known and unknown states, and the estimations of
transition probabilities for the set of known states $H\subseteq Q$,
we consider that the set of states $\hat{H}= H\times S $ in
$\overline{\calM}$ is known and construct a  sub-\ac{mdp}
$\overline{\calM}_{\hatH}$ of $\overline{\calM}$ that only includes
the set of known states $\hatH$, with an additional sink (or
absorbing) state that groups together the set of unknown states
$V\setminus \hatH$.  A policy is computed in order to maximize the
probability of hitting some target set in $\overline{\calM}_{\hatH}$
within a \emph{finite}-time horizon
$T$. 
We show that by following this policy, in $T$ steps, either there is a
high probability of hitting a state in the accepting end states of
$\calM$, or some unknown state will be explored, which at some point
will make an unknown state to be known.

Once all states become known, the structure of $M$ must have been
identified and the set of accepting end components in the learned
product \ac{mdp} $\overline{\calM}$ is exactly these in the true
product \ac{mdp} $\calM$. As a result, with probability at least
$1-\delta$, the policy obtained in $\overline{\calM}$ is \emph{near
  optimal}. Informally, a policy $f$ is near optimal, if, from any
initial state, the probability of satisfying the specification with
$f$ in $T$ steps is no less than the probability of eventually
satisfying the specification with the optimal policy, minus a small
quantity.
\paragraph*{Example} Consider the \ac{mdp} taken from
\citep[p.855]{Baire2004}, as a running example.  The objective is to
always eventually visiting the state $q_3$.  That is,
$\varphi=\square\lozenge q_3$. In \cite{Baire2004}, the \ac{mdp} is
fully known and the algorithm for computing the optimal policy is
given. As the \ac{mdp} has already encoded the information of the
specification, the atomic propositions are omitted and we can use the
\ac{mdp} $M$ as the product \ac{mdp} $\calM$ with acceptance condition
$\{(\emptyset, \{q_3\})\}$ and the accepting end component is
$(\{q_3\}, f(q_3)=\alpha)$.
 \begin{figure}[t]
\centering
 \includegraphics[width=0.5\textwidth]{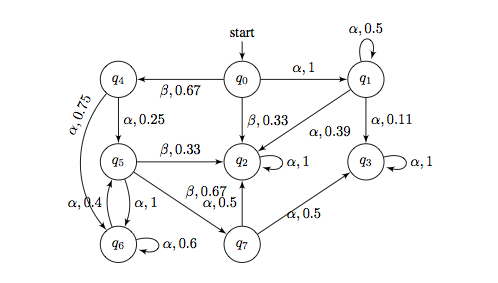}
 \caption{Example of an \ac{mdp} with states $Q=\{q_i, i =0,\ldots,
   7\}$, actions $\Sigma=\{\alpha, \beta\}$, and transition
   probability function $P$ as indicated.}
\label{fig:exmdp} 
\end{figure}
For this known \ac{mdp}, with respect to the specification
$\square\lozenge q_3$, the optimal policy $f^\ast$ and the probability
of satisfying the specification under $f^\ast$ is obtained in
Table~\ref{tb:ex-optimal}.
\begin{table}[H]
\centering
\caption{The  optimal policy and state values in the
  \ac{mdp} of Fig.~\ref{fig:exmdp}. \label{tb:ex-optimal}
}
\scalebox{0.9}{
\begin{tabular}{l c c c c c c c c}
\toprule
 {} &$q_0$ & $q_1$ & $q_2$ & $q_3$ & $q_4$ & $q_5$ & $q_6$ & $q_7$ \\
\hline
$f^\ast(\cdot)$ &$ \beta$ &$ \alpha$ &$ \alpha$ &$ \alpha$ & $ \alpha$ &$ \beta$& $
\alpha$ & $ \alpha$\\
$U_\calM^\ast(\cdot)$ &$0.22445$ & $0.22$
 & $0 $ & $1$ & $ 0.335$ & $ 0.335$ &$ 0.335 $&$ 0.5$ \\
\bottomrule
\end{tabular}}
\end{table}
\subsection{Maximum likelihood estimation of transition probabilities}
For the \ac{mdp} $M$, we assume that for each state-action pair, the
probability distribution $\mathrm{Dist}(q,a) : Q\rightarrow[0,1] $,
defined by $\mathrm{Dist}(q,a)(q') = P(q,a,q')$, is an independent
Dirichlet distribution (follows the assumptions in
\cite{duff2002optimal,Wang2005,CastroP07}).  For each $(q,a)\in
Q\times \Sigma$, we associate it at time $t$ for some $ t\ge 0$ with a
positive integer vector $\theta^t_{q,a} $, where $\theta^t_{q,a}(q')$
is the number of observations of transition $(q,a,q')$.  The agent's
\emph{belief} for the transition probabilities at time $t$ is denoted
as $\theta^t$ where $\theta^t= \{ \theta^t_{q,a},  (q,a)\in
Q\times \Sigma\}$. Given a transition $(q_1,\sigma,q_2)$, the belief
is updated by $\theta^{t+1}_{q,a}(q')= \theta^{t}_{q,a}(q')+1$ if
$q=q_1, a= \sigma, q' =q_2$, otherwise
$\theta^{t+1}_{q,a}(q')=\theta^{t}_{q,a}(q') $. Let
$\norm{\theta^t_{q,a} }_1= \sum_{q'\in Q} \theta^t_{q,a}(q')$.  

At time $t$, with $\theta_{q,\sigma}^t(q')$ large enough, the
\emph{maximum likelihood estimator} \cite{balakrishnan2004primer} of
the transition probability $P(q,\sigma,q')$ is a random variable of
normal distribution with mean  and variance, respectively,
\[
\overline{P}(q,\sigma,q') =\frac{ \theta_{q,\sigma}^t (q')
}{\norm{\theta^t_{q,\sigma}}_1}, 
\mathrm{Var}= \frac{
  \theta_{q,\sigma}^t (q')(\norm{ \theta_{q,\sigma}^t }_1 -
  \theta_{q,\sigma}^t (q'))}{\norm{ \theta_{q,\sigma}^t }_1^2(\norm{
    \theta_{q,\sigma}^t }_1 +1)}.
\]


\subsection{Approximating the underlying \ac{mdp}}
We extend the definition of $\alpha$-approximation in \ac{mdp}s  
\cite{Kearns2002}, to labeled \ac{mdp}s.
\begin{definition}
\label{def:approximation}
  Let $M$ and $\overline{M}$ be two labeled \ac{mdp}s over the
  \emph{same state and action spaces} and let $0< \alpha < 1$.  $\overline{M}$ is an
  $\alpha$-approximation of $M$ if $\overline{M}$ and $M$ share the
  same labeling function and the same structure, and for any state $q_1$
  and $q_2$, and any action $a \in \Sigma$, it holds that
$ \abs{ P(q_1,a,q_2)- \overline{P}(q_1,a,q_2)} \le \alpha .$
\end{definition}
By construction of the product \ac{mdp}, it is easy to prove that if
$\overline{M}$ $\alpha$-approximates $M$, then $\overline{\mathcal{M}}
= \overline{M} \ltimes \mathcal{A}_{\varphi}$ is an
$\alpha$-approximation of $\mathcal{M} = M\ltimes
\mathcal{A}_{\varphi}$.  In the following, we denote the true \ac{mdp}
(and its product \ac{mdp}) by $M$ (and $\calM$), the
learned \ac{mdp} (and the learned product \ac{mdp}) by
$\overline{M}$ (and $\overline{\calM}$).

In Problem~\ref{mainproblem}, since the true \ac{mdp} is unknown, at
each time instance, we can only compute a policy $f$ using our
hypothesis for the true model.  Thus, we need a method for evaluating
the performance of the synthesized policy. For this purpose, based on
the simulation lemma in \cite{Brafman2003,Kearns2002}, the following
lemma is derived. It provides a way of estimating the $T$-step state
values under the synthesized policy in the unknown \ac{mdp} $\cal M$,
using the \ac{mdp} learned from observations and the approximation
error between the true \ac{mdp} and our hypothesis.

\begin{lemma}
\label{lm:simulation}
Given two \ac{mdp}s $M = \langle Q,\Sigma, P, \calAP, L\rangle$ and
$\overline{M}=\langle Q, \Sigma,\overline{P}, \calAP,L \rangle$. If
$\overline{M}$ is an $\frac{\epsilon}{NT}$ -approximation of $M$ where
$N$ is the number of states in $M$ (and $\overline{M}$), $T$ is a
finite time horizon, and $0< \epsilon <1$, then for any specification
automaton $\mathcal{A}_{\varphi} =\langle S, 2^{\calAP}, T_s, I_s,
\mathsf{Acc} \rangle$, for any state $v$ in the product \ac{mdp}
$\calM= M\ltimes \mathcal{A}_{\varphi} =\langle V,\Sigma, \Delta, v_0,
\mathsf{Acc}\rangle$, for any policy $f: V\rightarrow \Sigma$, we have
that
$
\abs{U_{\mathcal{M}}^f(v, T)- U_{\overline {\cal{M}}}^f(v,T)} \le
\epsilon .
$ 
\end{lemma}
The proof is given in Appendix. It worths mentioning that though the
confidence level $1-\delta$ is achieved for the estimation of each
transition probability, the confidence level on the bound between
$U_{\mathcal{M}}^f(v, T)$ and $ U_{\overline {\cal{M}}}^f(v,T)$ for
$T$ steps is not $(1-\delta)^T$. The reader is referred to the proof
for more details.

Lemma \ref{lm:simulation} is important in two aspects. First, for any
policy, it allows to estimate the ranges of $T$-step state values in
the true \ac{mdp} using its approximation.  We will show in
Section~\ref{subsec:exploreexploit} that the learned \ac{mdp}
approximates the true \ac{mdp} for some $0< \alpha <1$.  Second, it
shows that for a given finite time horizon $T$, the size of the
specification automaton will not influence the accuracy requirement on
the learned \ac{mdp} for achieving an $\epsilon$-close $T$-step state
value for any policy and any initial state. Therefore, even if the
size of the specification automaton is exponential in the size of the
temporal logic specification, this exponential blow-up will not lead
to any exponential increase of the required number of samples for
achieving a desired approximation through learning. Yet, the
specification influences the choice of $T$ potentially. In the
following we will discuss how to choose such a finite time horizon $T$
and the potential influence.

\begin{lemma}
\label{lm:compare_opt_policy}
Let $\overline{M}$ be an $\frac{\epsilon}{NT}$-approximation of
$M$. For any specification automaton $\calA_\varphi$, suppose $f:
V \rightarrow \Sigma$ and $g: V\rightarrow \Sigma$ be
the $T$-step optimal policy in $\overline{\calM}=\overline{M}\ltimes
\calA_\varphi$ and $\calM=M\ltimes\calA_\varphi$ respectively. For any
state $v \in V$, it holds that
$
\abs{U_\calM^f(v, T) - U_{\calM}^{g}(v,T)} \le 2\epsilon.
$
\end{lemma}
\begin{proof}
  It directly follows from $U_{\overline{\calM}}^g(v,T) \le
  U_{\overline{\calM}}^f(v,T)$, $ \abs{U_\calM^f(v, T)
    -U_{\overline{\calM}}^f(v,T)} \le \epsilon$ and $
  \abs{U_{\overline{\calM}}^g(v,T) -U_\calM^g(v, T)} \le \epsilon$,
  which can be derived from Lemma~\ref{lm:simulation}.
\end{proof}

The finite time horizon $T$ is chosen in a way that for the optimal
policy $f$, the state-value $U^f_{\cal M}(v, T)$ has to be
sufficiently close to the  probability of satisfying the
specification eventually (an infinite horizon), that is, $U_{\cal
  M}^f(v)$.

\begin{definition}[$\epsilon$-state value mixing time]
  Given the product \ac{mdp} $\mathcal{M}$ and a policy $f$,
  let $d^f(t)=\max_{v\in V} \abs{ U_{\mathcal{M}}^f (v, t)-
    U_{\mathcal{M}}^f (v) }$, and the \emph{$\epsilon$-state value
    mixing time} is defined by
$
t_{\mix}^f(\epsilon):= \min \{t: \ d^f(t)\le \epsilon\}.
$
\end{definition}
Thus, given some $0 < \epsilon <1$, we can use an (estimated) upper
bound of the $\epsilon$-state value mixing time $t_{\mix}^f(\epsilon)$
for the optimal policy $f$ as the finit time horizon $T$.
\subsection{Exploration and exploitation}
\label{subsec:exploreexploit}
In this section, we use an exploration-exploitation strategy similar
to that of the R-max algorithm \cite{Brafman2003}, in which the choice
between exploration and exploitation is made implicit. The basic idea
is that the system always exercises a $T$-step optimal policy in some
\ac{mdp} constructed from its current knowledge (exploitation). Here
$T$ is chosen to be $\epsilon$-state value mixing time of the optimal
policy. It is guaranteed that if there exists any state for which the
system does not know enough due to insufficient observations, the
probability of hitting this unknown state is non-zero within $T$
steps, which encourages the agent to explore the unknown states. Once
all states are known, it is ensured that the structure of the
underlying \ac{mdp} has been identified. Then, based on
Lemma~\ref{lm:simulation} and \ref{lm:compare_opt_policy}, the
$T$-step optimal policy synthesized with our hypothesis performs
nearly as optimal as the true optimal policy.

We now formally introduce the notions of known states and known
\ac{mdp} following \cite{Kearns2002}.
\begin{definition}[Known states]
\label{def:knownstates}
Let $M$ be an \ac{mdp} and $\calA_\varphi$ be the specification
automaton. Let $q$ be a state of $M$ and $\sigma\in \Sigma$ be an
action enabled from $q$. Let $T$ be the $\epsilon$-state-value mixing
time of the optimal policy in $\calM=M\ltimes \calA_\varphi$.  A
probabilistic transition $(q,\sigma,q')$ is \emph{known} if with
probability at least $1-\delta$, we have for any $q'\in Q$, $
\mathrm{Var} \cdot k \le \frac{\epsilon}{NT} $, where $k$ is the
critical value for the $1-\delta$ confidence interval, $\mathrm{Var}$
is the variance of the maximum likelihood estimator for the transition
probability $P(q,\sigma,q')$, $N$ is the number of states in $M$.  A
state $q$ is \emph{known} if and only if for any action $\sigma$
enabled from $q$, and for any state $q'$ that can be reached by action
$\sigma$, the probabilistic transition $(q,\sigma,q')$ is
known.  \end{definition}


\begin{definition}
  \label{def:submdp}Given $ H \subseteq Q$  the set of known states
  in an \ac{mdp} $M$, let $\hatH\times S \subseteq V$ be the set of
  known states in the product \ac{mdp} $\calM$.  The \emph{known}
  product \ac{mdp} is $\calM_{\hatH} = \langle \hatH\cup \{\sink\},
  \Sigma, \Delta_{\hatH}, v_0, \mathsf{Acc}_{\hatH} \rangle $ where
  $\hatH\cup \{\sink\}$ is the set of states and $\sink$ is the
  absorbing/sink state. $\Delta_{\hatH}$ is the transition probability
  function and is defined as follows: If both $v,v'\in \hatH$,
  $\Delta_{\hatH}(v,\sigma,v')=\Delta(v,\sigma,v')$. Else if $v\in
  \hatH$ and there exists $\sigma \in \Sigma$ such that $\Delta(v,
  \sigma,v')>0$ for some $v'\notin \hatH$, then let
  $\Delta_{\hatH}(v,\sigma,\sink)= \sum_{v'\notin \hatH}
  \Delta(v,\sigma,v')$. For any $\sigma \in \Sigma$,
  $\Delta_{\hatH}(\sink,\sigma,\sink)=1$.  The acceptance condition $
  \mathsf{Acc}_{\hatH} $ in $\calM_{\hatH}$ is a set of pairs
  $(\{(\hat{J}_i \cap \hatH, \hat{K}_i \cap \hatH ) \mid
  i=0,\ldots,m\} \cup \{(\emptyset, \{\sink\}) \} )\setminus
  \{(\emptyset,\emptyset)\} $.
\end{definition}
Intuitively, by including $(\emptyset, \{\sink\})$ in
$\mathsf{Acc}_{\hatH}$, we encourage the exploration of unknown
states  aggregated in $\sink$.

\paragraph*{Example (cont.)}
In the example \ac{mdp}, we treat $M$ as the product \ac{mdp} $\calM$.
Initially, all states in \ac{mdp} (Fig.~\ref{fig:exmdp}) are unknown,
and thus the known product \ac{mdp} has only state $\sink$, see
Fig.~\ref{fig:initial}.  Figure~\ref{fig:submdp} shows the known product
\ac{mdp} $M_H$ where $H=\{q_2,q_3,q_5,q_6\}$.
\begin{figure}[t]
\centering
\includegraphics[width=0.5\textwidth]{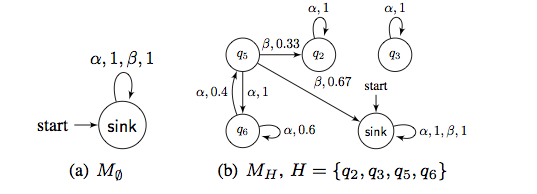}
\caption{Two known product \ac{mdp}s constructed from the example
  \ac{mdp} (the same as the product \ac{mdp})
  with the sets of known states $\emptyset$ and $ \{q_2,q_3,q_5,q_6\}$
  respectively.}
\end{figure}

The following lemma shows that the optimal $T$-step policy in
$\mathcal{M}_{\hatH}$ either will be near optimal in the product
\ac{mdp} $\mathcal{M}$, or will allow a rapid exploration of an
unknown state in $M$.

\begin{lemma}
\label{lm:exploitexplore}
Given a product \ac{mdp} $\calM$ and a set of known states
$\hatH\subset V$, for any $v\in \hatH$, for $0< \alpha <1$, let $f$ be the optimal
$T$-step policy in $\mathcal{M}_{\hatH}$. Then, one of the following
two statements holds: \begin{inparaenum}[1)]
  \item $U_{{\cal M}}^f(v, T) \ge U_{\cal M}^{\ast}(v, T) -\alpha $;
    \item An unknown state which is not in the accepting end state set $\calC$ will be visited in the course of running
    $f$ for $T$ steps with a probability at least $\alpha$.
\end{inparaenum}
\end{lemma}
\begin{proof}
  Suppose that $U_{\cal M}^f(v, T)< U_{\cal M}^{\ast}(v, T) -\alpha$
  (otherwise, $f$ witnesses the claim).  First, we show for any policy $g: V\rightarrow \Sigma$, for any
  $v\in \hat{H}$, it holds that
\begin{equation}
\label{eq:inequality}
U_{\calM_{\hatH}}^g(v,T) \ge U_{\calM}^g(v,T)
\end{equation}
To prove this, for notation simplicity, let $\Pr=\Pr_{\calM^g}$ be the probability
  measure over the paths in $\calM^g$ and $\Pr' =
  \Pr_{\calM_{\hat{H}}^g}$ be the probability measure over the paths
  in $\calM_{\hat{H}}^g$.

Let $X \subseteq V^\ast$ be a
  set of paths in $\calM^g$ such that each $x \in X$, with $\abs{x}\le
  T$, starts in $ v$, ends in $\calC$ and has every state in $\hatH$;
  $Y \subseteq V^\ast$ be the set of paths in $\calM^g$ such that each
  $y \in Y$, with $\abs{y}\le T$, starts in $v$, ends in $\calC$ and
  has at least one state not in $\hatH$; and $Y'$ be the set of paths
  $y$ in $\calM^g_{\hatH}$ which starts in $v$, ends with $\sink $ and
  has length $\abs{y}\le T$.  We can write
\begin{align*}
 &U_{\cal M}^g(v, T)= \sum_{x\in X} \Pr(x)  + \sum_{y \in Y} \Pr(y)
, \text{ and }\\
 &U_{{\cal M}_{\hatH}}^g(v, T)= \sum_{x \in X} \Pr'(x)  + \sum_{y \in Y'}
 \Pr'(y).
\end{align*}
Since the transition probabilities in $\mathcal{M}$ and
$\mathcal{M}_{\hatH}$ are same for the set of known states, and $X$ is
the set of paths which only visit known states, we infer that $
\sum_{x \in X} \Pr(x) = \sum_{x \in X} \Pr'(x)$. Moreover, since $y\in
Y$ contains an unknown state, it leads to $\sink$ in $\calM_{\hatH}$,
and thus is in $ Y'$. We infer that $Y \subseteq Y'$, $ \sum_{y \in Y}
\Pr(y) \le \sum_{y \in Y'} \Pr'(y) $ and thus $U_{{\cal
    M}_{\hatH}}^g(v, T)\ge U_{\cal M}^g(v,
T)$. 

Next, let $f$ be the optimal $T$-step policy in $\calM_{\hatH}$ and $\ell$ be the
optimal $T$-step policy in $\calM$. From Eq.~\eqref{eq:inequality}, we
obtain an inequality: $U_{\calM_{\hatH}}^\ell(v,T) \ge U_{\calM}^\ell (v,T)$.

By the $T$-step optimality of $f$ in $\calM_{\hatH}$ and $\ell$ in
$\calM$, it  also holds that  $ U_{\calM_{\hatH}}^f(v,T) \ge
U_{\calM_{\hatH}}^\ell(v,T)$ and $ U_{\calM}^\ast (v,T) = U_{\calM}^\ell(v,T) \ge
U_{\calM}^f(v,T)$. Hence,
\begin{multline*}
U_{\calM_{\hatH}}^f(v,T) \ge 
U_{\calM_{\hatH}}^\ell(v,T) \ge U_{\calM}^\ast (v,T) \ge
U_{\calM}^f(v,T) \\
\implies
U_{\calM_{\hatH}}^f(v,T) -  U_{\calM}^f(v,T) \ge U_{\calM}^\ast (v,T)
- U_{\calM}^f (v,T).
\end{multline*}
Given the fact that $U_{\calM}^\ast (v,T) - U_{\calM}^f (v,T) >
\alpha$, we infer that
\[
U_{{\cal M}_{\hatH}}^{f}(v,T) -U_{\cal M}^{f}(v,T)= \sum_{z \in Z}
\Pr(z) > \alpha ,
\]
where $Z$ is the set of paths such that each $z\in Z$ with $\abs{z}\le
T$, $z$ starts from $v$, and ends in some \emph{unknown} state which
is \emph{not} an accepting end state in $\calM$.  Therefore, we reach at the
conclusion that if $U_{\calM}^f(v,T) < U^\ast_{\calM}(v,T)-\alpha$,
then the probability of visiting an unknown state which is not in $\calC$ must be at least
$\alpha$.
\end{proof}
Note that, for any unknown state which is in $\calC$, one can apply
the policy in its corresponding accepting end component to visit such
a state infinitely often, and after a sufficient number of visits, it
will become known.

Though we use the product \ac{mdp} $\calM$ in
Lemma~\ref{lm:exploitexplore},  Lemma~\ref{lm:exploitexplore} can also
be applied
to the learned product \ac{mdp} $\overline{\calM}$.

\section{\ac{pac-mdp} algorithm in control with temporal logic constraints.}
\label{sec:algorithm}

\begin{theorem}
\label{thm1}

Let $M=\langle Q,\Sigma, P,\calAP, L\rangle$ be an \ac{mdp} with $P$
unknown, for which we aim to maximize the probability of satisfying a
\ac{ltl} formula $\varphi$.  Let $0 < \delta < 1$, and $\epsilon >0$
be input parameters. Let $\calM=
M\ltimes \calA_\varphi$ be the product \ac{mdp} and $T$ be the
$\epsilon$-state value mixing time of the optimal policy in
$\calM$. Let $F_{\cal M}(\epsilon,T)$ be the set of policies in $\cal
M$ whose $\epsilon$-state value mixing time is $T$. With probability
no less than $1-\delta$, Algorithm~\ref{alg:learnandsynthesize} will
return a policy $f \in F_{\cal M}(\epsilon, T)$ such that
$\abs{U^f_{\calM} (v, T) - U^\ast_{\calM}(v) } \le 3\epsilon $ within
a number of steps polynomial in $\abs{Q}$, $\abs{\Sigma}$, $\abs{S}$, $T$,
$\frac{1}{\epsilon}$ and $\frac{1}{\delta}$.
\end{theorem}
\begin{proof} Firstly, applying the Chernoff bound
  \cite{mark2011probability}, the upper bound on the number of visits
  to a state for it to be known is polynomial in $\abs{\Sigma}, T,
  \frac{1}{\epsilon}$ and $\frac{1}{\delta}$. Before all states are
  known, the current policy $f$ exercised by the system is $T$-step
  optimal in $\overline{\calM}_{\hatH}$ induced from the set of known
  states $\hatH$. Then, by Lemma~\ref{lm:exploitexplore}, either for
  each state, policy $f$ attains a state value $\alpha$-close to the
  optimal $T$-step state value in $\overline{\calM}$, or an unknown
  state will be visited with probability at least $\alpha$. However,
  because $\overline{\calM}_{\hatH}$ is
  $\frac{\epsilon}{NT}$-approximation of $\calM_{\hatH}$,
  Lemma~\ref{lm:simulation} and Lemma~\ref{lm:compare_opt_policy}
  guarantee that policy $f$ either attains a state value
  $(2\epsilon+\alpha)$-close to the optimal $T$-step state value in
  $\calM$ for any state, or explores efficiently. If it is always not
  the first case, then after a finite number of steps, which is
  polynomial in $\abs{Q},\abs{\Sigma}, T, \frac{1}{\epsilon},
  \frac{1}{\delta}$, all states will be known, and the learned
  \ac{mdp} $\overline{M}$ (resp. $\overline{\calM}$)
  $\frac{\epsilon}{NT}$-approximates the true \ac{mdp} $M$
  (resp. $\calM$).  Since $T$ is the $\epsilon$-state value mixing
  time of the optimal policy in $\calM$, the $T$-step optimal policy
  $g:V\rightarrow \Sigma$ in $\calM$ satisfies $\abs{U_{\calM}^g(v,T)
    -U_\calM^\ast(v)} \le \epsilon$. From
  Lemma~\ref{lm:compare_opt_policy}, it holds that
  $\abs{U_{\calM}^f(v,T) - U_{\calM}^g(v,T)}\le 2\epsilon$ and thus we
  infer that $\abs{U_{\calM}^f(v,T) -U_\calM^\ast(v)} \le 3\epsilon $.
\end{proof}
Note that, the sample complexity of the algorithm is polynomial in
$\abs{Q},\abs{\Sigma}, T, \frac{1}{\epsilon}, \frac{1}{\delta}$,
unrelated to the size of $\calA_\varphi$. However, in the value
iteration step, the space and time complexity of policy synthesis is
polynomial in $\abs{Q}, \abs{S}, \abs{\Sigma}, T, \frac{1}{\epsilon},
\frac{1}{\delta}$.

In problem~\ref{mainproblem}, we aim to obtain a policy $f$ which is
$\varepsilon$-optimal in $\calM$. This can be
achieved by setting $\epsilon=\frac{\varepsilon}{3}$ (see
Theorem~\ref{thm1}).

In Algorithm~\ref{alg:learnandsynthesize}, policy is updated at most
$\abs{Q}$ times as there is no need to update it if a new observation does not cause an unknown state to
become known.  Given the fact that for \ac{ltl} specifications, the
time and space complexity of synthesis is polynomial in the
size of the product \ac{mdp}, Algorithm~\ref{alg:learnandsynthesize}
is a provably efficient algorithm for learning and policy
update.

Similar to \cite{Kearns2002,Brafman2003}, the input $T$ can be
eliminated by either estimating an upper bound for $T$ or starting
with $T=1$ and iteratively increase $T$ by $1$. The reader can refer
to \cite{Kearns2002,Brafman2003} for more detailed discussion on the
elimination technique.

\begin{algorithm}[h]
  \SetKwFunction{KnownMDP}{KnownMDP}
  \SetKwFunction{ValueIteration}{ValueIteration}
  \SetKwFunction{Exploit}{Exploit}
  \SetKwFunction{BalancedWandering}{BalancedWandering}
  \SetKwFunction{Update}{Update} { \KwIn{The state and action sets $Q$
      ,$\Sigma$, the set of atomic propositions $\calAP$ and the
      labeling function $L: Q\rightarrow
      2^{\calAP}$, the specification \ac{dra} $\calA_\varphi$,
      parameters $\epsilon$ and $\delta$, the (upper bound of)
      $\epsilon$-state mixing time $T$ for the optimal policy in
      $M\ltimes \calA_{\varphi}$. } \KwOut{A policy $f:Q\times
      S\rightarrow \Sigma$.}
    \Begin { $H : = \emptyset$, $q:= q_0$, $s:= I_s$, \textsf{recompute}=True;

$\overline{M}=\langle Q,\Sigma, \overline{P}, \calAP, L\rangle$,
\tcc{$\overline{P}(q,a,q')=0$ for any $(q,a,q')\in Q\times
\Sigma\times Q $.}

\While{True}{ $s:=T_s(s, L(q))$,

\If{\textsf{recompute}=True}{
 $\hatH=H\times S$, $\overline{\calM}
  = \overline{M}\ltimes \calA_{\varphi}$, $\overline{\calM}_{\hatH}: =
  \KnownMDP(\overline{\calM}, \hatH)$, $
  f:=\ValueIteration(\overline{\calM}_{\hatH}, T)$}

$q' ,a=
  \Exploit((q,s),f)$,
$H_p=H$,
$\overline{M}, H: = \Update(\overline{M},H, q, a, q', \epsilon,
\delta, \abs{Q}, T)$

\lIf{$H_p \ne
H$}{\textsf{recompute}=True}\lElse{\textsf{recompute}=False}
\lIf{$\overline{P}(q',a,q')=1$ or $(q,s) \in
  \overline{\mathcal{C}}$}{With probability $0\le p\le1$, restart with
  a random state in $ Q$ and set $s=I_s$, else $q:=q'$}\lElse{ $q :
  =q'$}  \tcc*{$\overline{\calC}$ is the accepting end states in
  $\overline{M}$.}

\If{$H = Q$}{
\Return{$f:=\ValueIteration(\overline{\calM}, T).$}}
}}}
\caption{\texttt{LearnAndSynthesis}
\label{alg:learnandsynthesize}}
\end{algorithm}
During learning, it is possible that for state $q \in Q$ and for
action $a\in \Sigma$, we estimate that $\overline{P}(q,a,q)=1$. Then
either in the true \ac{mdp}, $P(q,a,q)=1$, or, $P(q,a,q) <1$ yet we
have not observed a transition $(q,a,q')$ for any $q'\neq q$. In this
case, with some probability $p$, we restart with a random initial
state of \ac{mdp}. With probability $1-p$, we keep exploring state
$q$. The probability $p$ is a tuning parameter in
Algorithm~\ref{alg:learnandsynthesize}.

\section{Examples}
\label{sec:examples}
We apply Algorithm~\ref{alg:learnandsynthesize} to the running example \ac{mdp}
(Fig.~\ref{fig:exmdp}) and a robotic motion planning
problem in an unknown terrain. The implementations are in Python on a
desktop with Intel(R) Core(TM) processor and $16$ GB of memory.

\begin{figure*}[ht]
\centering
\subfigure[]{
\includegraphics[width=0.3\textwidth]{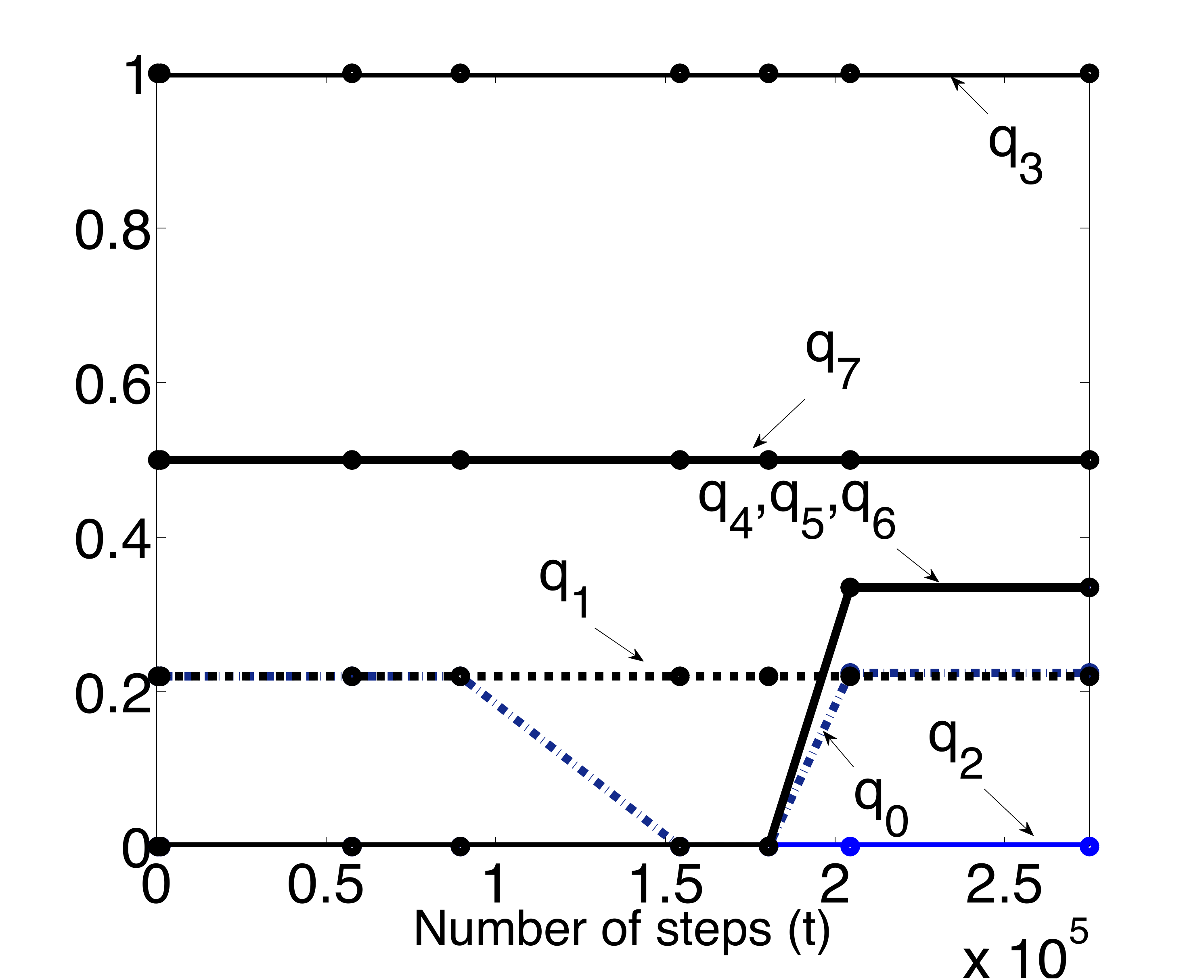}
\label{fig:learningmdp}
}
\subfigure[]{
\includegraphics[width=0.3\textwidth]{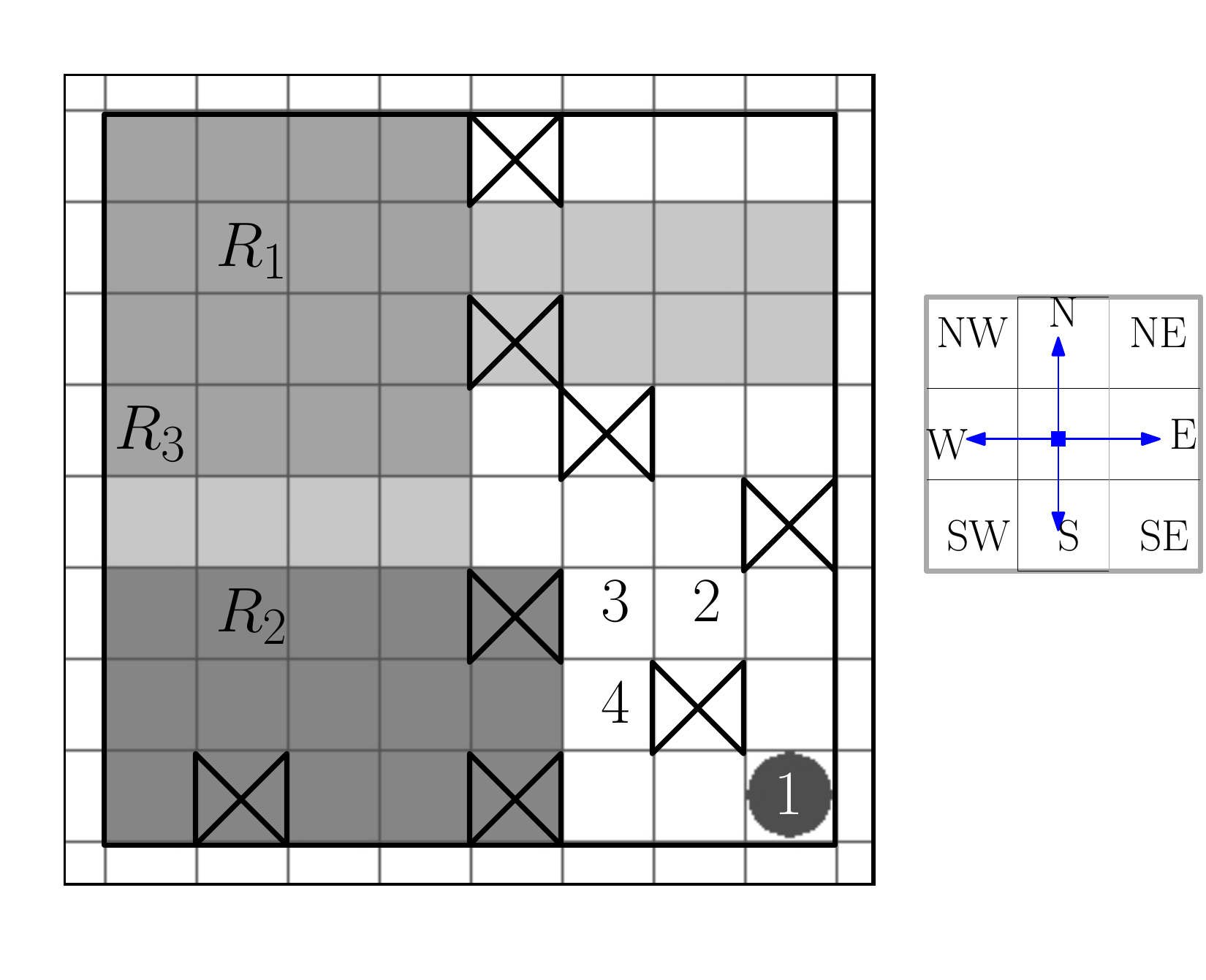}
\label{gridworldimg}
}
\subfigure[]{
\includegraphics[width=0.3\textwidth]{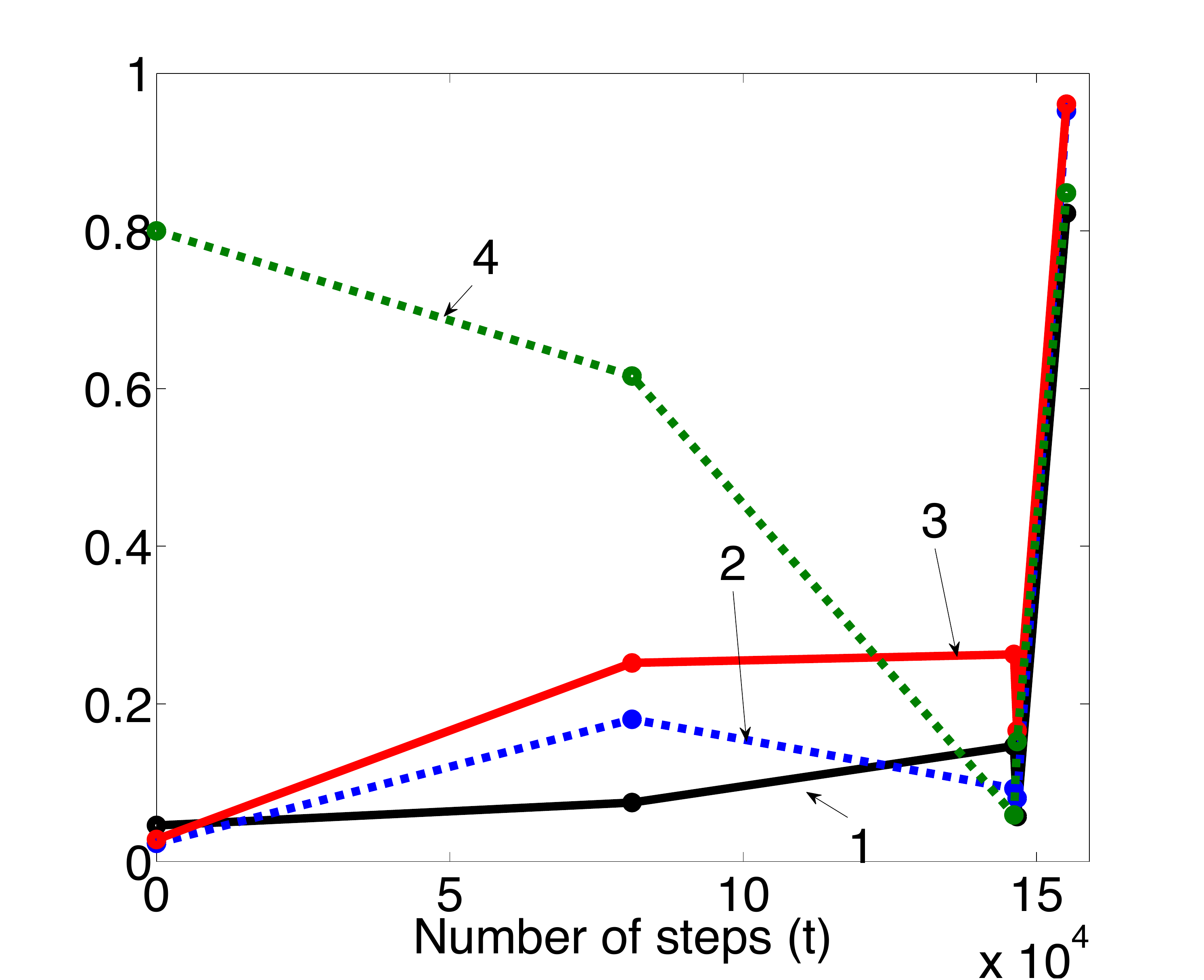}
\label{fig:gridworldevaluate}}
\vspace{-2ex}
\caption{(a) The state value $U_{\calM}^{f^t}(q_i)$, $i=0,\ldots, 7$
  v.s.  step $t$, with $\varepsilon=0.01$, $\delta=0.05$ and
  $T=15$. The markers represents the steps when the policy is
  recomputed. Note that, for states $q_0$ and $q_1$, the state
  values  at $t_f$ are $0.22445$ and $0.22$ respectively, which are
  indiscernible from the figure.
(b) Left: A $10 \times 10$ gridworld, where the disk
  represents the robot, the cells $R_1$, $R_2$, and $R_3$ are the
  interested regions, the crossed cells are the obstacles, labeled
  with $R_4$, the cells on the edges are walls, and we assume that if
  the robot hits the wall, it will be bounced back to the previous
  cell. Different grey scales represents different terrains: From the
  darkest to the lightest, these are ``grass,'' `` pavement,''
  ``sand'' and ``gravel.'' Right: The transitions of the robot, in
  which the center cell is the current location of the robot. (c) The
  state value $U_{\calM}^{f^t}((q_0,s_0))$ v.s.  step $t$, where $q_0
  \in \{1,2,3,4\}$ is the initial cell and $s_0=T_s(I_s,L(q_0))$,
  under $\epsilon=0.01$, $\delta=0.05$ and $T=50$. The markers
  represents the steps when the policy is recomputed.}
\label{fig:gridworldevaluate}
\end{figure*}
\subsection{The running example}

We consider different assignments for $\varepsilon$ and $95\%$ of
confidence level, i.e., $\delta=0.05$, and $T=15$ as the (estimated
upper bound of) $\frac{\varepsilon}{3}$-state value mixing time of the
optimal policy, for all assignments of $\varepsilon$. A \emph{step}
means that the system takes an action and arrives at a new state.

For $\varepsilon=0.01$, after $t_f=274968$ steps, all states become
known and the policy update terminates after $8$ updates and $35.12$
seconds. Let the policy at step $t$ be denoted as $f^t$. We evaluate
the policy $f^t$ in the true product-\ac{mdp} and plot the state value
(the probability of satisfying the specification from that state under
policy $f^t$) $U_{\calM}^{f^t}(q_i): i =0,\ldots, 7$, for the finite
horizon $[0,t_f]$ in Fig.~\ref{fig:learningmdp}. Note that, even
though the policy computed at step $t=204468$ has already converged to
the optimal policy, it is only after $t=t_f=274968$ that in the system's
hypothesis, the $T$-step state value computed using the known \ac{mdp}
with its $T$-step optimal policy converges $\varepsilon$-close to the
optimal state value in the true product \ac{mdp}.


For $\varepsilon=0.02$, in $136403$ steps with $17.73$ seconds, all
states become known and the policy is the optimal
one. 
For $\varepsilon=0.05$, in $55321$ steps with $7.18 $ seconds all
states are known.  However, the policy $f$ outputs $\alpha$ for all
states except $q_5$, at which it outputs $\beta$. Comparing to the
optimal policy which outputs $\beta$ for both $q_0$ and $q_5$, $f$ is
sub-optimal in the true \ac{mdp} $\calM$: With $f$,
$U_\calM^f(q_0)=0.22$, comparing to $0.22445$ with the optimal
policy. For the remaining states, we obtain the same state values with
policy $f$ as the optimal one.


Finally, in three experiments, it is observed that the actual maximal
error ($0.00445$ with $\varepsilon=0.05$) never exceeds $0.01$,
because we use the loose upper bound on the error between the $T$-step
state value with any policy in $\calM$ and its approximation
$\overline{\calM}$ in Lemma~\ref{lm:simulation}, to guarantee the
correctness of the solution.


\subsection{A motion planning example}

We apply the algorithm to a robot motion planning problem (see
Fig.~\ref{gridworldimg}). The environment consists of four different
unknown terrains: Pavement, grass, gravel and sand. In each terrain
and for robot's different action (heading north (`N'), south (`S'),
west (`W') and east (`E')), the probability of arriving at the correct
cell is in certain ranges: $[0.9,0.95] $ for pavement, $[0.85,0.9]$
for grass, $[0.8,0.85]$ for gravel and $[0.75,0.80]$ for sand. With a
relatively small probability, the robot will arrive at the cell
adjacent to the intended one. For example, with action `N', the
intended cell is the one to the north (`N'), whose the adjacent ones
are the northeast (`NE')and northwest cells (`NW') (see
Fig.~\ref{gridworldimg}).
The objective of the robot is to maximize the probability of
satisfying a temporal logic specification $\varphi= \square \lozenge
(R_1 \land \lozenge (R_2 \land \lozenge R_3)) \land \square \neg R_4 $
where $R_1,R_2,R_3$ are critical surveillance cells and $R_4$ includes
a set of unsafe cells to be avoided.  For illustrating the
effectiveness of the algorithm, we mark a subset of cells labeled by
$1,2,3,4$ and evaluate the performance of iteratively updated policies
given that a cell in the set is the initial location of the robot.

Given $\varepsilon=0.01$, $\delta=0.05$, and $T=50$, all states become
known in $t_f=155089$ steps and $1593.45$ seconds, and the policy
updated four times (one for each terrain type).  It is worth
mentioning that most of the computation time is spent on computing the
set of bottom strongly connected components using the algorithm in
\cite{Chatterjee2012} in the structure of the learned \ac{mdp}, which
is then used to determine the set of accepting end components in
$\overline{\calM}$.  In Fig.~\ref{fig:gridworldevaluate}, we plot the
state value $U_{\calM}^{f^t}((q_0,s_0))$ where $q_0\in \{1,2,3,4\}$
and $s_0=T_s(I_s,q_0)$ for a finite time horizon. The policy output by
Algorithm~\ref{alg:learnandsynthesize} is the optimal policy in the
true \ac{mdp}. The video demonstration for this example is available
at \url{http://goo.gl/rVMkrT}.
\section{Conclusion and future work} 
\label{sec:conclusion}
We presented a \ac{pac-mdp} method for synthesis with temporal
logic constraints in unknown MDPs and developed an algorithm that
integrates learning and control for obtaining approximately
optimal policies for temporal logic constraints with polynomial time,
space and sample complexity.  Our current work focuses on other
examples (e.g. multi-vehicle motion
planning),   comparison to alternative, possibly ad hoc methods,
and implementing a version of Algorithm~\ref{alg:learnandsynthesize}
that circumvents the need for the input $T$ following
\cite{Brafman2003}.

There are a number of interesting future extensions. First, although
here we only considered one-player stochastic games, it is also
possible to extend to two-player stochastic games, similar to the
R-max algorithm \cite{Brafman2003}.  The challenge is that in
two-player stochastic games, only considering deterministic,
memoryless policies in the product-\ac{mdp} may not be sufficient. For
the synthesis algorithm, different strategy classes (memoryless,
finite-memory, combined with deterministic and randomized) require
different synthesis methods. We may need to integrate this
\ac{pac-mdp} approach with other synthesis methods for randomized,
finite-memory strategies.
Second, besides the objective of maximizing the
probability of satisfying a temporal logic constraint, other
objectives can be considered, for example, minimizing the weighted
average costs \cite{wolff2012optimal}. Third, the method is
model-based in the sense that a hypothesis for the underlying
\ac{mdp} is maintained. The advantage in such a model-based approach is that
when the control objective is changed, the knowledge gained in the
past can be re-used in the policy synthesis for the new
objective. However, model-free \ac{pac-mdp} approach
\cite{strehl2006pac}, in which information on the policy is retained
directly instead of the transition
probabilities, can be of interests as
its space-complexity is asymptotically less than the space
requirement for model-based approaches.

\appendix


\begin{proof}[Proof of Lemma \ref{lm:simulation}]
  By Definition~\ref{def:approximation}, $M$ and $\overline{M}$ share
  the same structure. Thus, for any \ac{dra} $\mathcal{A}_\varphi$,
  the product \ac{mdp}s $\calM=M\ltimes \mathcal{A}_\varphi$ and
  $\overline{\calM}= \overline{M}\ltimes \mathcal{A}_{\varphi}$ share
  the same structure and the same set of accepting end states
  $\calC\subseteq V$.

  For any policy $f$, let $M_i$ be the Markov chains
  obtained from the induced Markov chains $\mathcal{M}^f$ and
  $\overline{\mathcal{M}}^f$ in the following way: Start at $v$ and for the first $i$
  transitions, the transition probabilities are the same as in
  $\overline{\mathcal{M}}^f$, and for the rest of steps, the transition
  probabilities are the same as in $\mathcal{M}^f$.  Clearly,
  $\mathcal{M}^f =M_0$ and $\overline{\mathcal{M}}^f=M_T$.  For
  notational simplicity, we denote $ h_{M_i}(\cdot)=h_i (\cdot) $,
  $\Pr_{M_i}(\cdot)=\Pr_i(\cdot)$.  Then, we have that
\begin{align}
\small
  &\abs{U_{\cal M}^f(v, T) - U_{\overline{\cal M}}^f (v, T)} =
  \abs{h_0^{\le T}(v, \calC) -h_T^{\le T} (v, \calC) } \notag \\ 
  = & \left| h_0^{\le T}(v, \calC) -h_1^{\le T} (v, \calC)+h_1^{\le
      T}(v, \calC) -h_2^{\le T} (v, \calC) +\ldots +  \right. \notag\\
& \left.h_{T-1}^{\le
      T}(v, \calC) -h_T^{\le T} (v, \calC)
  \right| \notag= \sum_{i=0}^{T-1} \abs{ h_{i}^{\le T}(v, \calC) -h_{i+1}^{\le T}
    (v, \calC)}   \notag \\
\le & T \cdot \max_{i\in \{0,\ldots T-1\}}\abs{
    h_{i}^{\le T}(v, \calC) -h_{i+1}^{\le T} (v, \calC)}. \label{eq0}
\end{align}
For any $i=0,\ldots, T-1$, we have that
\begin{multline*}
\small
  \mathsf{Diff1}= \abs{h_{i}^{\le T}(v, \calC)
    -h_{i+1}^{\le T}(v, \calC)} = \left| h_{i}^{\le i}(v,\calC)
  \right.\\
\left.+
    \sum_{k=i+1}^T h_i^{k}(v,\calC) - h_{i+1}^{\le i}(v,\calC)
    -\sum_{k=i+1}^T h_{i+1}^{k}(v,\calC) \right|.
\end{multline*}
Since for the first $i$ steps, the transition probabilities in $M_i$
and $M_{i+1}$ are the same, then the probabilities of hitting the set
$\calC$ in $M_i$ and $M_{i+1}$ equal to the probability of hitting
$\calC$ in $\calM^f=M_0$, i.e., $h_{i}^{\le i}(v,\calC)=h_{i+1}^{\le
  i}(v,\calC) =h_0^{\le i}(v,\calC)$.  Remind that $ \Pr_i(x)$, for
some $x \in V^\ast$, is the probability of path $x$ occurring in $M_i$,
as a consequence,
\begin{align*}
\small
\begin{split}
&\mathsf{Diff1}= \abs{\sum_{k=i+1}^T
h_i^{k}(v,\calC) - \sum_{k=i+1}^T
h_{i+1}^{k}(v,\calC)}\\
= &\left| \sum_{v' \notin \calC} \Pr_i(x v') \sum_{v'' \notin
  \calC} \big(\Pr_i(v'v'')  
\cdot h_i^{\le T-i-1}(v'',\calC) \big) - \right. \\
 &\left. \sum_{v'  \notin \calC} \Pr_{i+1}(x v') \sum_{v'' \ \notin
  \calC} \big(  \Pr_{i+1}(v'v'')  
 \cdot h_{i+1}^{\le T-i-1}(v'',\calC) \big)\right|,
\end{split}
\end{align*}
where $x \in V^\ast$ is a path of length $i-1$ that starts in $v$ and
does not contain any state in $\calC$. Note that for the first $i$ (resp. the last $T-i-1$)
transitions, the transition probabilities in $M_i$ and $M_{i+1}$ are
the same as these in $\calM^f = M_0$ (resp. $\overline{\calM}^f
=M_T$), thus we have $ \Pr _i(x v') = \Pr_{i+1}(x v')=
\Pr_0(x v')$ and $h_{i}^{\le T-i-1}( v'',\calC) = h_{i+1}^{\le
  T-i-1}( v'',\calC) =h_T^{\le T-i-1}( v'',\calC)$ . 

Let $v'=(q',s')$, $v''=(q'',s'')$ and $a=f(v')$. It is also noted that
$ \Pr_i(v'v'') = \Pr_i((q',s')(q'',s''))=P(q',a, q'')$ with $s''=
T_s(s', L(q''))$ and $ \Pr_{i+1}(v'v'') = \Pr_{i+1}((q',s')(q'',s''))=
\overline{P}(q',a,q'')$.  Thus,  as $\overline{M}$ approximate $M$, we have
\begin{align*}
\small
\begin{split}
   &\mathsf{Diff1}= \sum_{v' \notin \calC} \Pr_0(x v') \sum_{v'' \notin
  \calC}  \big(\abs{ P(q',a, q'')- \overline{P}(q',a, q'')
} \\
&  \cdot h_T^{\le T-i-1} (v'', \calC)\big) \\
\le &\sum_{v' \notin \calC} \Pr_0(x v')\cdot \frac{ \epsilon}{NT} \cdot
\sum_{v'' \notin \calC} h_T^{\le T-i-1} (v'', \calC) =\mathsf{Diff2},
\end{split}
\end{align*}
The first term $\sum_{v' \notin \calC} \Pr_0(xv')\le 1 $ and the last
term is the sum of the probabilities of visiting $\calC$ from
different states in $ V\setminus \calC$ within $T-i-1$ steps, each of
which is bounded by $1$. Moreover, since $v''=(q'',s'')$ where $s''$
is determined by the previous state $s'$ in $\calA_\varphi$ and the
current state $q''$, i.e., $s''=T_s(s',L(q''))$, the sum is bounded by
the number $N$ of states (choices for $q''$) in the underlying
\ac{mdp} $M$. Thus, $ \mathsf{Diff1}\le \mathsf{Diff2}\le \frac{
  \epsilon}{NT} \cdot N = \frac{\epsilon}{T}$.  Finally, from
\eqref{eq0}, we have $\abs{U_{\cal M}^f(v, T) - U_{\overline{\cal
      M}}^f (v, T)} \le \frac{ \epsilon}{T} \cdot T = \epsilon$.
\end{proof}
\bibliographystyle{ieeetr}

\bibliography{refer}

\begin{thebibliography}{10}

\bibitem{Kearns2002}
M.~Kearns and S.~Singh, ``{Near-optimal reinforcement learning in polynomial
  time},'' {\em Machine Learning}, vol.~49, pp.~209--232, Nov. 2002.

\bibitem{Brafman2003}
R.~Brafman and M.~Tennenholtz, ``{{R-MAX}-a general polynomial time algorithm
  for near-optimal reinforcement learning},'' {\em The Journal of Machine
  Learning}, vol.~3, pp.~213--231, 2003.

\bibitem{legay2010statistical}
A.~Legay, B.~Delahaye, and S.~Bensalem, ``{Statistical model checking: An
  overview},'' in {\em Runtime Verification}, pp.~122--135, Springer, 2010.

\bibitem{Henriques2012}
D.~Henriques, J.~G. Martins, P.~Zuliani, A.~Platzer, and E.~M. Clarke,
  ``{Statistical model checking for Markov decision processes},'' {\em
  International Conference on Quantitative Evaluation of Systems}, pp.~84--93,
  Sept. 2012.

\bibitem{Chen2012}
Y.~Chen, J.~Tumova, and C.~Belta, ``{LTL robot motion control based on automata
  learning of environmental dynamics},'' in {\em IEEE International Conference
  on Robotics and Automation}, pp.~5177--5182, May 2012.

\bibitem{FuAtAlCDC2013}
J.~Fu, H.~G. Tanner, and J.~Heinz, ``Adaptive planning in unknown environments
  using grammatical inference,'' in {\em IEEE Conference on Decision and
  Control}, 2013.

\bibitem{abs-1212-3873}
H.~Mao, Y.~Chen, M.~Jaeger, T.~D. Nielsen, K.~G. Larsen, and B.~Nielsen,
  ``Learning markov decision processes for model checking,'' in {\em
  Proceedings of Quantities in Formal Methods} (U.~Fahrenberg, A.~Legay, and
  C.~R. Thrane, eds.), Electronic Proceedings in Theoretical Computer Science,
  pp.~49--63, 2012.

\bibitem{rutten2004mathematical}
J.~Rutten, M.~Kwiatkowska, G.~Norman, and D.~Parker, {\em Mathematical
  Techniques for Analyzing Concurrent and Probabilistic Systems, {\rm P.
  Panangaden and F. van Breugel (eds.)}}, vol.~23 of {\em CRM Monograph
  Series}.
\newblock American Mathematical Society, 2004.

\bibitem{Baire2004}
C.~Baier, M.~Gr{\"o\ss}er, M.~Leucker, B.~Bollig, and F.~Ciesinski,
  ``{Controller Synthesis for Probabilistic Systems (Extended Abstract)},'' in
  {\em Exploring New Frontiers of Theoretical Informatics} (J.-J. Levy,
  E.~Mayr, and J.~Mitchell, eds.), vol.~155 of {\em International Federation
  for Information Processing}, pp.~493--506, Springer US, 2004.

\bibitem{bianco1995model}
A.~Bianco and L.~{De Alfaro}, ``{Model checking of probabilistic and
  nondeterministic systems},'' in {\em Foundations of Software Technology and
  Theoretical Computer Science}, pp.~499--513, Springer, 1995.

\bibitem{de1997formal}
L.~{De Alfaro}, {\em {Formal Verification of Probabilistic Systems}}.
\newblock PhD thesis, Stanford University, 1997.

\bibitem{Chatterjee2012}
K.~Chatterjee, M.~Henzinger, M.~Joglekar, and N.~Shah, ``{Symbolic algorithms
  for qualitative analysis of Markov decision processes with B\"{u}chi
  objectives},'' {\em Formal Methods in System Design}, vol.~42, no.~3,
  pp.~301--327, 2012.

\bibitem{duff2002optimal}
M.~O. Duff, {\em Optimal Learning: Computational Procedures for Bayes-adaptive
  Markov Decision Processes}.
\newblock PhD thesis, University of Massachusetts Amherst, 2002.

\bibitem{Wang2005}
T.~Wang, D.~Lizotte, M.~Bowling, and D.~Schuurmans, ``Bayesian sparse sampling
  for on-line reward optimization,'' in {\em Proceedings of the 22nd
  International Conference on Machine Learning}, pp.~956--963, ACM, 2005.

\bibitem{CastroP07}
P.~S. Castro and D.~Precup, ``Using linear programming for bayesian exploration
  in markov decision processes,'' in {\em International Joint Conferences on
  Artificial Intelligence} (M.~M. Veloso, ed.), pp.~2437--2442, 2007.

\bibitem{balakrishnan2004primer}
N.~Balakrishnan and V.~B. Nevzorov, {\em A Primer on Statistical
  Distributions}.
\newblock Wiley, 2004.

\bibitem{mark2011probability}
B.~L. Mark and W.~Turin, {\em Probability, Random Processes, and Statistical
  Analysis}.
\newblock Cambridge University Press Textbooks, 2011.

\bibitem{wolff2012optimal}
E.~M. Wolff, U.~Topcu, and R.~M. Murray, ``Optimal control with weighted
  average costs and temporal logic specifications.,'' in {\em Robotics: Science
  and Systems}, 2012.

\bibitem{strehl2006pac}
A.~L. Strehl, L.~Li, E.~Wiewiora, J.~Langford, and M.~L. Littman, ``{PAC}
  model-free reinforcement learning,'' in {\em Proceedings of the 23rd
  International Conference on Machine Learning}, pp.~881--888, ACM, 2006.

\end{thebibliography}

\end{document}